\documentclass[journal]{IEEEtran}
\usepackage[cmex10]{amsmath}
\usepackage{amssymb}
\usepackage{cite}
\usepackage{graphicx}

\usepackage{algorithmic}

\usepackage{setspace}

\usepackage{array}

\usepackage{fixltx2e}

\usepackage{color}

\usepackage{times,color}

\newcommand{\mb}{\mathbb}

\newtheorem{theorem}{Theorem}

\newtheorem{algorithm}[theorem]{Algorithm}

\begin{document}
\title{Protograph-Based Raptor-Like LDPC Codes }

\author{Tsung-Yi Chen,~\IEEEmembership{Member,~IEEE,} 
Kasra Vakilinia,~\IEEEmembership{Student Member,~IEEE,}
Dariush Divsalar,~\IEEEmembership{Fellow,~IEEE,} and Richard D. Wesel,~\IEEEmembership{Senior~Member,~IEEE}\\
tsungyi.chen@northwestern.edu, vakiliniak@ucla.edu, Dariush.Divsalar@jpl.nasa.gov, wesel@ee.ucla.edu
\thanks{R.~D.~Wesel, K. Vakilinia and D.~Divsalar are with the Department of Electrical Engineering, University of California, Los Angeles, Los Angeles, CA 90095, USA. T.-Y. Chen is with the Electrical Engineering and Computer Science Department, Northwestern University, Evanston, IL 60208, USA}%
\thanks{This material is based upon work supported by the National Science Foundation under Grant Number 1162501. Any opinions, findings, and conclusions or recommendations expressed in this material are those of the author(s) and do not necessarily reflect the views of the National Science Foundation. This research was carried out in part at the Jet Propulsion Laboratory, California Institute of Technology, under a contract with NASA. Parts of this work were presented at the Global Communications Conference 2011 and the International Conference on Communications 2012. }
}

\maketitle

\begin{abstract}
\boldmath
This paper proposes a class of rate-compatible LDPC codes, called protograph-based Raptor-like (PBRL) codes. The construction is focused on binary codes for BI-AWGN channels. As with the Raptor codes, additional parity bits are produced by exclusive-OR operations on the precoded bits, providing extensive rate compatibility. Unlike Raptor codes, the structure of each additional parity bit in the protograph is explicitly designed through density evolution.  The construction method provides low iterative decoding thresholds and the lifted codes result in excellent error rate performance for long-blocklength PBRL codes. For short-blocklength PBRL codes the protograph design and lifting must avoid undesired graphical structures such as trapping sets and absorbing sets while also seeking to minimize the density evolution threshold. Simulation results are shown in information block sizes of $k=192$, $16368$ and $16384$. Comparing at the same information block size of $k=16368$ bits, the PBRL codes outperform the best known standardized code, the AR4JA codes in the waterfall region. The PBRL codes also perform comparably to DVB-S2 codes even though the DVB-S2 codes use LDPC codes with longer blocklengths and are concatenated with outer BCH codes. 
\end{abstract}
\begin{IEEEkeywords}
Channel coding, Low-Density Parity-Check Codes.
\end{IEEEkeywords}

\section{Introduction}
\label{sec:Intro}
\IEEEPARstart INCREMENTAL redundancy (IR) systems with receiver confirmation are widely used in modern communication systems, e.g. in the 3GPP-LTE standard. Receiver confirmation refers to a class of feedback systems where the confirmation (a decision to conclude a transmission session) is determined at the receiver. For general discussion of feedback systems and the various types of confirmation see \cite{Chen_feedback_2013}. 

To achieve high expected throughput, modern IR systems often use a family of good rate-compatible channel codes that provides improved error protection as the number of received symbols available to the decoder increases.  This paper provides a general technique for constructing families of rate-compatible low-density parity-check (LDPC) codes and provides numerical results showing excellent performance.

The remainder of this section briefly reviews previous work on the design of rate-compatible channel codes and then provides a summary of the main contributions of this paper. 

\subsection{Rate-Compatible Channel Codes}

Rate-compatible punctured convolutional (RCPC) codes and rate-compatible punctured turbo (RCPT) codes are among the most popular rate-compatible channel codes used in IR systems. 
For both RCPC and RCPT codes, a collection of rate-compatible puncturing patterns are often carefully designed to ensure good error rate performance across the family of rates despite the rate-compatible constraint.  In \cite{Hagenauer_Rate_1988}, Hagenauer optimized performance when the puncturing patterns are restricted to be periodic with a relatively short period. Following Hagenauer's framework, Rowitch et al. studied RCPT codes with periodic puncturing patterns \cite{Rowitch_performance_2000}. Analyzing the performance of randomly punctured turbo codes, Liu and Soljanin showed that RCPT code performance degrades significantly when the punctured code rate is above a threshold \cite{Liu_punctured_2003}.

Low-Density Parity-Check (LDPC) codes were first introduced by Gallager in his dissertation in 1963 \cite{Gallager_1963_Low}. Gallager defines an $(n, d_v, d_c)$ LDPC code as a length-$n$ binary code with a parity-check matrix containing $d_v$ ones in each column and $d_c$ ones in each row. These LDPC codes are now referred to as regular LDPC codes.

Tanner \cite{Tanner_recursive_1981} proposed the construction of a class of long-blocklength codes that combines a collection of short blocklength codes through a bipartite graph. He also generalized the decoding algorithm proposed by Gallager. Tanner's work introduced the representation of LDPC codes as bipartite graphs. 
MacKay et al. \cite{MacKay_1999_Good} showed that LDPC codes provide capacity-approaching performance similar to turbo codes \cite{Berrou_1993_Near} when decoded by a message-passing algorithm with soft information. MacKay et al. also proposed several heuristics to construct good LDPC codes. 

In contrast to regular LDPC codes, irregular LDPC codes have parity-check matrices that have a variety of column weights and row weights.  By optimizing the variable-node and check-node degree distributions, Luby et al. \cite{Luby_2001_Improved} showed that properly constructed irregular LDPC codes can achieve rates even closer to capacity than the regular ones. Richardson, Shokrollahi and Urbanke \cite{Richardson_2001_Design} created a systematic method called density evolution to design and analyze the optimal degree distribution of LDPC codes based on the assumption that the blocklength can be infinitely long.

Because of their capacity-approaching performance at individual rates,  LDPC codes provide a promising structure for constructing a family of rate-compatible codes. Aiming to achieve high throughput in IR systems for various classes of channels, numerous heuristics have been proposed to construct rate-compatible LDPC codes. The first work in the construction of rate-compatible LDPC codes appears to be \cite{Li_2002}. See also \cite{Yazdani_construction_2004, Jacobsen_2007, El-Khamy_design_2009} and the references therein.

Ha et al. studied the asymptotic behavior of rate-compatible punctured LDPC codes based on density evolution \cite{Ha_rate_2004}. They also studied the design of puncturing patterns for LDPC codes with relatively short blocklengths \cite{Ha_rate_2006}. Many heuristics have been proposed for designing better puncturing patterns to enhance the error-rate performance and/or to allow efficient encoding. See for example, \cite{Song_linear_2008, Kim_deisgn_2009, Vellambi_finite_2009}.

The analysis of IR systems using rate-compatible LDPC codes has been studied by Sesia et al. \cite{Sesia_incremental_2004} based on random coding and density evolution of infinite-length LDPC ensemble. The focus in \cite{Sesia_incremental_2004} is on long-blocklength codes for wireless fading channels whereas this paper focuses on memoryless channels and studies both short-blocklength and long-blocklength regimes. 

Although not rate-compatible,  a row-combining approach is used in \cite{Casado_multiple_2009} to support a variety of rates that share the same encoder structure and have the same blocklength.  

Thorpe \cite{Thorpe_2003} introduced a new class of LDPC codes called protograph-based LDPC codes, or protograph codes.  These codes were studied extensively by Divsalar et al. \cite{Divsalar_2009}. The design of protograph codes begins with the construction of a relatively small bipartite graph called the protograph. After using density evolution to properly design the protograph , the protograph is copied many times and the edges are permuted carefully to obtain a bipartite graph with a desired blocklength.  As discussed in \cite{Divsalar_2009}, protograph codes allow efficient decoder implementation in hardware.

Obtaining a family of rate-compatible LDPC codes through rate-compatible puncturing of a low-rate mother code is straightforward. However, it is commonly observed that punctured finite-length LDPC codes suffer from a larger performance degradation compared to punctured turbo codes at high rates \cite{Yazdani_construction_2004}. Another way to construct rate-compatible codes is by the method of extending codes. Yazdani et al. studied the construction of rate-compatible LDPC codes based on a combination of extending and puncturing \cite{Yazdani_construction_2004}. Yazdani et al. concluded that a combination of the two methods yields better rate-compatible codes than using puncturing alone. The current paper constructs a family of rate-compatible LDPC codes by similar techniques. Our focus, however, is on a special class of protograph codes that provides numerous benefits in encoding and decoding complexity while also achieving excellent performance across a variety of rates.

The design of rate-compatible protograph codes first appeared in \cite{Dolinar_rate_2005}. Nguyen et al. further optimized the design of rate-compatible protograph code families by extrinsic information transfer (EXIT) analysis and a greedy search of a well-chosen collection of nested protographs \cite{Nguyen_design_2012}. Using density evolution, the current paper also studies the construction of a family of rate-compatible protograph codes by extending high-rate protograph codes. In contrast to \cite{Dolinar_rate_2005} and \cite{Nguyen_design_2012}, this paper focuses on the design of protograph code families that have a similar structure to the class of rateless codes called Raptor codes \cite{Shokrollahi_2006_Raptor}. Constraining our design to the structure of Raptor codes makes the construction and optimization manageable while providing outstanding performance and extensive rate-compatibility.  

Following the Raptor-like structure proposed in \cite{Chen_2011_Globecom}, Nitzold et al. applied  spatial coupling \cite{Nitzold_spatially_2012} to improve the threshold, which can be also viewed as Raptor-like LDPC convolutional codes \cite{Felstrom_time_1999}. Nitzold et al. focused on the analysis of the asymptotic decoding threshold where the rate loss due to the time-spreading number $L$ is negligible. In addition to asymptotic threshold analysis, this paper also studies the construction of finite-length LDPC codes.

A recent work by Nguyen et al. \cite{Nguyen_rate_2013} considers a general structure for extending rate-compatible protograph codes but ends up proposing an example protograph that has the Raptor-like structure as first proposed in \cite{Chen_2011_Globecom} and \cite{Chen_2012_ICC}.

\subsection{Main Contributions and Organization}

This paper proposes a class of rate-compatible LDPC codes called protograph-based raptor-like (PBRL) LDPC codes. The construction and optimization of PBRL codes are discussed and simulation results are presented. Comparing to existing codes in the literature (e.g. AR4JA codes in \cite{CCSDS_ORANGE}, DVB-S2 codes in \cite{DVBS2} and protograph codes in \cite{Nguyen_design_2012}), PBRL codes show outstanding performance while providing extensive rate-compatibility.

The rest of the paper is organized as follows: Sec.~\ref{sec:PBRL} presents the construction of PBRL codes and Sec.~\ref{sec:Optimize} provides the optimization methods for designing PBRL codes. Examples of constructing PBRL codes are given in Sec.~\ref{sec:ExamplesPBRL}.  Finally Sec.~\ref{sec:ConclusionPBRL} concludes the paper.

\section{Protograph-Based Raptor-Like LDPC Code}
\label{sec:PBRL}
This section introduces the structure of PBRL codes. The encoding and decoding of PBRL codes are also discussed. 

\subsection{Overview of Protograph Codes and Raptor Codes}\label{sec:ProtographOverview}

In order to facilitate efficient hardware implementation of the decoder, LDPC codes are often constructed with certain structures, e.g. quasi-cyclic LDPC codes and regular LDPC codes \cite{BookLin}. Protograph-based LDPC codes, or simply protograph codes, are also structured codes and allow efficient hardware implementation \cite{Divsalar_2009}. 

\begin{figure}[t]
\centering
\includegraphics[width=0.45\textwidth]{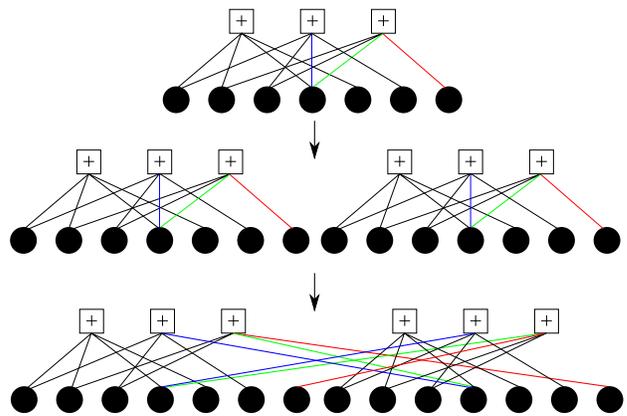}
\caption{Lifting process of the $[7, 4]$ Hamming code. }
\label{fig:HammingLifting}
\end{figure}

A protograph, or projected graph, is a bipartite graph with a relatively small number nodes. A copy-and-permute operation, often referred to as ``lifting'', is applied to the protograph to obtain larger graphs of various sizes, resulting in longer-blocklenth LDPC codes.  As part of lifting, the variable-node connections of the edges of the same type are permuted among the protograph replicas.

An $[n, k]$ code is a linear block code that encodes $k$ bits into $n$ bits. Fig.~\ref{fig:HammingLifting} shows an example of the lifting process for the $[7,4]$ Hamming code. For clarity only three types of edges are permuted and each type is color coded with the same color. Note that the protograph can also have parallel edges, i.e., multiple edges connecting a pair of variable node and check node. Parallel edges in a protograph can later be removed through the lifting process. A detailed discussion on parallel edges in protographs can be found in \cite{Divsalar_2009}.

Introduced by Luby \cite{Luby_2002} and Shokrollahi \cite{Shokrollahi_2006_Raptor} respectively, LT codes and Raptor codes share many similarities with LDPC codes and are shown to achieve the binary erasure channel (BEC) capacity universally. Etesami et al. \cite{Etesami_2006} explored the application of Raptor codes to binary memoryless symmetric channels and derive various results, including the fact that Raptor codes are not universal except for the BEC. Note that results on Raptor codes such as \cite{Shokrollahi_2006_Raptor} and \cite{Etesami_2006} rely heavily on the assumption of large information blocks.

\subsection{The Structure of PBRL Codes}\label{sec:PBRLOverview}
\begin{figure}[t]
\centering
\includegraphics[width=0.4\textwidth]{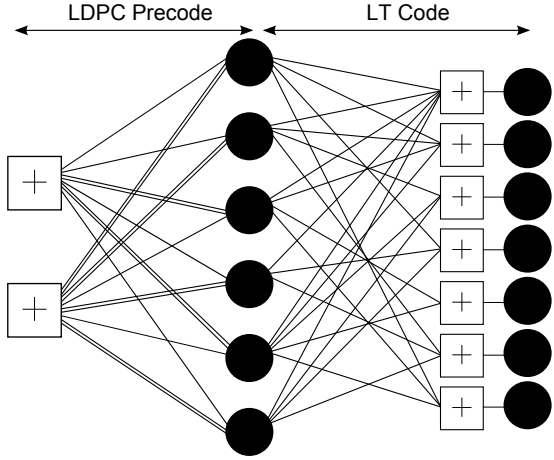}
\caption{Protograph for a PBRL code with a rate-$2/3$ precode.}
\label{fig:r34_ProtoRaptor_PBRL}
\end{figure}	
The structure of a PBRL code can be best illustrated by its protograph. Fig.~\ref{fig:r34_ProtoRaptor_PBRL} shows the protograph of a PBRL code. This protograph consists of two parts: $(1)$ a relatively simple protograph code (on the left) representing the protograph of the precode and $(2)$ a number of check nodes (on the right) that are each connected to several variable nodes of the precode and one additional variable node that has only a single edge. The second part represents the protograph of an LT code. The highest rate shown in the example of Fig.~\ref{fig:r34_ProtoRaptor_PBRL} is $2/3$, which is obtained by transmitting only the variable nodes of the precode.  Lower-rates are obtained by transmitting the variable nodes in the LT-code protograph starting from the top node.

After the lifting operation, the first part can be seen as an LDPC precode in a Raptor code, and the degree-one variable nodes of the second part can be efficiently encoded with the precoded symbols in a manner similar to the LT code. Hence the structure of this protograph code resembles a Raptor code, but with a deterministic (rather than random) encoding rule for combining the precoded symbols.

The bipartite graph of a protograph can be described by a protomatrix, which is the parity check matrix of a protograph. Let $\mathbf{0}$ be the all zero matrix and $\mathbf{I}$ be the identity matrix with the appropriate dimensions. The protomatrix of the protograph shown in Fig.~\ref{fig:r34_ProtoRaptor_PBRL} is given as
\begin{align}
\label{eqn:protomatrix_r34_eg}
H = \left[
\begin{array}{cc}
H_p   &\mathbf{0}\\
H_{LT}&\mathbf{I}
\end{array} 
\right]
\end{align}
where $H_{\text{p}}$ and $H_{\text{LT}}$ are given as
\begin{align}
\label{eqn:protomatrix_r34_2}
H_{\text{p}} = \left[\begin{array}{cccccc}
	1 & 1 & 2 & 1 & 2 & 1 \\
	2 & 2 & 1 & 2 & 1 & 2
	\end{array} \right]
\end{align}
and
\begin{align}
H_{\text{LT}} = \label{eqn:protomatrix_r34_3}
\left[\begin{array}{ccccccc}
	 1 & 1 & 1 & 1 & 1 & 1 \\
	 1 & 1 & 1 & 0 & 1 & 0 \\
	 0 & 1 & 0 & 0 & 1 & 1 \\
	 1 & 0 & 0 & 1 & 0 & 1 \\
	 0 & 0 & 1 & 0 & 1 & 0 \\
	 0 & 1 & 0 & 1 & 0 & 1 \\
	 1 & 0 & 1 & 0 & 1 & 0 
	\end{array} \right].
\end{align}
Hence to fully express the protomatrix of a PBRL protograph it is enough to specify $H_{\text{p}}$ and $H_{\text{LT}}$. Due to space limitation we will present the proposed protographs in terms of protomatrices whenever the structure of the protograph is clear. 

\subsection{Decoding and Encoding of PBRL Codes}
\label{sec:EncDec}
Consider the decoding of a traditional Raptor code that collects the precoded symbols and encodes them with an LT code. In the case of an LDPC precode used with an LT code, the decoding often proceeds as follows: the decoder first performs BP decoding on the LT code and then performs BP decoding on the precode. The two-stage decoding implies the use of two different BP decoders, each exchanging their extrinsic information after the iterative decoding.

In \cite{Soljanin_2006}, the authors note that because of the two-stage decoding, the complexity of Raptor codes is higher than that of rate-compatible LDPC codes. The PBRL code family always transmits the output symbols of the precode and has deterministic connections in the LT code.  These two properties facilitate joint decoding of the LT code part and the precode.  For traditional Raptor codes that use randomized encoding, the initial transmission of the LT symbols may not contain enough information for BP decoding to succeed even in a noiseless setting.  Always transmitting the precode symbols allows PBRL codes to have the potential for successful decoding after the initial transmission. 

For high-rate PBRL codes, the decoder can deactivate those check nodes in the LT part for which the neighboring degree-one variable node is not transmitted, implying implementation advantages. Thus, at the highest rate when only the precode is transmitted, none of the check nodes in the LT code part need to be activated, offering significant complexity reduction.

The encoding of the PBRL codes is as efficient as Raptor codes: after encoding the precode, the encoding of the LT code part only involves exclusive-or operations on the precode output symbols. For efficient encoding of the precode, see the discussion in \cite{Divsalar_2009} on efficient encoding of protograph codes.

Many techniques for extending LDPC codes exist in the literature as summarized in Sec.~\ref{sec:Intro}.  In comparison to the approaches in the literature, the Raptor-Like structure is very restrictive.  One might expect the structural constraints to also constrain performance as compared to less-restrictive structures for extending LDPC codes. One of the main conclusions of our paper is that despite the structural constraints, we obtain Raptor-like protographs with very low iterative decoding thresholds. By careful design of the protograph and the lifting process, the resulting finite-length codes can outperform existing rate-compatible LDPC codes that have been designed without the constraint of a Raptor-like structure. This observation provides a new perspective on designing rate-compatible LDPC codes that sometimes less is more.

\section{Optimization of Protograph-Based Raptor-Like LDPC Codes}
\label{sec:Optimize}
This section presents optimization procedures for finding good PBRL codes with short and long blocklengths. Belief propagation (BP) decoding is assumed. The optimization criteria are primarily based on minimizing the iterative decoding threshold. To simplify the computation we use a modified version of the reciprocal channel approximation (RCA) algorithm to compute the threshold. The following subsection briefly reviews the RCA. After presenting the modified RCA, we describe the optimization procedures and discuss the construction of protographs for PBRL codes.
\subsection{Density Evolution with Reciprocal Channel Approximation}
\label{sec:DensityEvo}
For BI-AWGN channels, the asymptotic \textit{iterative decoding threshold} \cite{Richardson_2001_BPCapacity} characterizes the performance of the ensemble of LDPC codes based on a specified protograph. This threshold indicates the minimum SNR required to transmit reliably with the underlying ensemble of codes as the blocklength grows to infinity.

Computing the exact iterative decoding threshold for BI-AWGN requires a large amount of computation. The RCA \cite{Chung_Dissertation}\cite{Divsalar_2009} provides a fast and accurate approximation to the density evolution originally proposed by Richardson et al. \cite{Richardson_2001_BPCapacity}\cite{Richardson_2001_Design}. Experimental results \cite{Divsalar_2009},\cite{Chung_Dissertation} show that the deviation from the exact density evolution is less than $0.01$ dB.

The RCA for BI-AWGN channel uses a single real-valued parameter $s$, the SNR, to approximate the density evolution. Define the reciprocal SNR as $r\in \mathbb{R}$ such that $C(s) + C(r) = 1$ where $C(s)$ is the capacity of the BI-AWGN channel with SNR $s$:
\begin{align}
C(s) = 1 - \int_{-\infty}^{\infty}
\log_2\left(1+ e^{-(2\sqrt{2s}u + 2s)}\right)\frac{e^{-u^2}}{\sqrt{\pi}}du .
\end{align}
The self-inverting reciprocal energy function \cite{Chung_Dissertation}
\begin{equation}
R(s) = C^{-1}\left(1-C(s)\right)
\end{equation}
transforms parameters $s$ and $r$ to each other. In other words, $r = R(s)$ and $s = R(r)$.

Let $s_{\text{chl}}$ be the channel SNR, $s_{e}$ be the message passed along an edge $e$ from a variable node to a check node and $r_{e}$ be the message passed along an edge $e$ from a check node to a variable node. Let $E_c$ be the set of edges that connect to a check node $c$ and $E_v$ be the set of edges that connect to a variable node $v$. 

RCA first initializes the message $s_e$ to $0$ if the edge $e$ is connected to a punctured variable node and to $s_{\text{chl}}$ otherwise. For all edges $e$ in the graph, RCA then computes a sequence of messages $(s_e^{(n)}, r_e^{(n)}), ~n = 0, \dots, N$ where $N$ is the maximum number of iterations. 

The original density evolution \cite{Richardson_2001_BPCapacity} determines the threshold based on the densities of all outgoing messages from variable nodes. Aiming to approximate this density evolution, RCA \cite{Chung_Dissertation} determines the decoding threshold $s_{\text{th}}$ as the minimum $s_{\text{chl}}$ such that $s_e^{(N)} > T$ for all edges $e$ in the graph, where $T$ is  a stopping threshold. 

Note that for the edge connecting to a degree-one variable node, $s_e = s_{\text{chl}}$ regardless of the number of iterations.  For this reason, the original RCA does not work if the graph contains degree-one variable nodes. We use a slightly modified version of the RCA that focuses on the overall reliability of each variable node $S_v$, rather than the reliability of every edge $s_e$.  This modification allows computation of a meaningful decoding threshold for protographs with degree-one variable nodes. Letting $N$ be the maximum number of iterations and $T>0$ be the stopping threshold, the modified RCA is summarized as follows:
\begin{algorithm}[Modified Reciprocal Channel Approximation]
Let $f_{\text{RCA}}(s_{\text{chl}}) : \mb{R}\mapsto\{0,1\}$ be a binary-valued function that returns $1$ if $s_{\text{chl}}$ is higher than $s_{\text{th}}$ and $0$ otherwise. To determine its output, the modified RCA computes the sequence $(s_e^{(n)}, r_e^{(n)}), n = 0, \dots, N$, for all edges $e$ in the graph. The computation of the sequences is given as follows:

	\begin{enumerate}
	\item[0)] For edges $e$ connected to punctured variable nodes, set $s_{e}^{(0)}=0$. For all other edges set $s_{e}^{(0)}=s_{\text{chl}}$.
	\item[1)] Generate the sequence $(s_e^{(n)}, r_e^{(n)})$ iteratively as follows: 
	\begin{align}
	r_e^{(n+1)} &= \sum\limits_{i\in E_c \setminus e} R\left(s_i^{(n)}\right)\,,
\\
	s_e^{(n+1)} &= s_{e}^{(0)} + \sum\limits_{i\in E_v \setminus e} R\left(r_i^{(n)}\right)\,.
 	\end{align}
 	\item[2)] For all variable nodes $v$ in the graph, compute $S_v^{(n)}$ as
 	\begin{equation}
 	S_v^{(n)} = S_v^{(0)} + \sum_{e\in E_v} s_e^{(n)}\,,
 	\end{equation}
 	where $S_v^{(0)}$ is defined as 
\begin{align}
S_v^{(0)} = 
\begin{cases}
0 & \text{if $v$ is punctured},
\\
s_{\text{chl}} & \text{otherwise}.
\end{cases}
\end{align}
	\item[3)] Let $S^* = \min\left\{S_v^{(N)}: \forall v \text{ in the graph}\right\}$. If $S^* > T$, $f_{\text{RCA}}(s_{\text{chl}})  = 1$. Otherwise $f_{\text{RCA}}(s_{\text{chl}})  = 0$.
	\end{enumerate}
\end{algorithm}
	Note that the values $s_e^{(n)}$ are additive at the variable nodes and the values $r_e^{(n)}$ are additive at the check nodes for all edges. 
	
	By the monotonicity of the threshold \cite{Richardson_modern_2008} we can perform bisection search at the desired level of precision using the function $f_{\text{RCA}}$. To increase the computation speed, a lookup table is used for computing $C(s)$ and $C^{-1}(s)$ and linear interpolation is used. 

\subsection{Optimizing the Precode Protograph of PBRL Codes} \label{sec:precode}
The design of the precode protograph follows the work of Divsalar et al. \cite[Sec.III]{Divsalar_2009}. One main conclusion of \cite{Divsalar_2009} is that code ensembles with a minimum variable node degree of $3$ or higher are guaranteed to have linear growth of their minimum distance with the blocklength. However, Divsalar et al. noted that judiciously adding some variable nodes with degree $2$ or even degree $1$ improves the threshold. 

The LT part of PBRL code familes inherently includes degree-one variable nodes. Therefore, the design of the precode protographs constrains all variable nodes in the precode to have degree at least three, ensuring the linear minimum distance growth property for the precode.  The structures we selected in our examples are inspired by the protograph examples in \cite{Divsalar_2009}, with some additional optimization through density evolution analysis and LDPC code simulation.

\subsection{Optimizing the LT Code Protograph of PBRL Codes}\label{sec:LTcode}

Given a precode protograph constructed based on the techniques described in \cite{Divsalar_2009}, we focus on the construction for the LT code part. The optimization algorithm of the LT code part is summarized as follows:
\begin{algorithm}
\label{OptProcedure}
Given a good protograph precode \cite{Divsalar_2009}, the construction of the LT code part for a PBRL code is given by the following steps:

	\begin{enumerate}
	\item Add a new check node and a new degree-one variable node connected to the new check node.
	\item Use the modified RCA to find the connections between the new check node and the precoded symbols that gives the lowest threshold.  
	\item If the lowest rate desired has been obtained, continue to step 4.  Otherwise, go to step 1.
	\item Lift the resulting protograph with circulant permutations to match the desired blocklengths for the rate-compatible family. The selection of the circulant permutations is based on the circulant progressive edge growth algorithm (cPEG) \cite{Hu_2005, Andrews_2004}.
	\end{enumerate}	
\end{algorithm}

As we will see later, parallel edges in the LT part can facilitate very low thresholds.  However, for short-blocklength codes these parallel edges introduce short cycles in the lifting process.  We observed empirically that for good performance both in terms of waterfall and error floor, parallel edges in the \textit{LT code} part of the protograph should be kept to a minimum (at most one pair of parallel edges in our examples) for short blocklength codes. 

Table~\ref{Threshold_PBRL} shows the thresholds of an optimized protograph for a short-blocklength PBRL code. The edge connections are give by replacing the following matrices in to \eqref{eqn:protomatrix_r34_eg}: the precode protomatrix is given as
\begin{align}
\label{eqn:protomatrix_r34_1}
H_{\text{p}} = \left[\begin{array}{cccccccc}
	4 & 1 & 1 & 2 & 1 & 2 & 1 & 2\\
	1 & 2 & 2 & 1 & 2 & 1 & 2 & 1 
	\end{array} \right],
\end{align}
and LT protomatrix is given as
\begin{align}
H_{\text{LT}} = \label{eqn:protomatrix_r34_2}
\left[\begin{array}{cccccccc}
	1 & 1 & 1 & 1 & 1 & 1 & 1 & 1\\
	1 & 1 & 1 & 1 & 1 & 1 & 1 & 0\\
	1 & 0 & 1 & 0 & 0 & 1 & 1 & 1\\
	0 & 1 & 0 & 0 & 1 & 0 & 1 & 1\\
	1 & 0 & 0 & 1 & 0 & 1 & 0 & 0\\
	0 & 0 & 1 & 0 & 1 & 0 & 1 & 0\\
	0 & 1 & 0 & 1 & 0 & 1 & 0 & 0\\
	1 & 0 & 0 & 0 & 0 & 0 & 0 & 1\\
	0 & 0 & 0 & 1 & 0 & 0 & 1 & 0
	\end{array} \right].
\end{align}

\begin{table}[t] 
	\caption{Thresholds of a PBRL code family of ($E_b/N_0$ in decibels). }	
	\label{Threshold_PBRL}
	\centering 
	\begin{tabular}{c|ccc} 
 	Rate & Threshold   & Shannon Limit & Gap \\  
	\hline 
	$6/8$   & 2.196 & 1.626  & 0.570\\ 
	$6/9$   & 1.804 & 1.059  & 0.745\\
	$6/10$  & 1.600 & 0.679  & 0.921\\
	$6/11$  & 1.464 & 0.401  & 1.063\\
	$6/12$  & 1.358 & 0.187  & 1.171\\
	$6/13$  & 1.250 & 0.018 & 1.232\\
	$6/14$  & 1.136 & -0.122 & 1.258\\
	$6/15$  & 1.016 & -0.238 & 1.254\\
	$6/16$  & 0.922 & -0.337 & 1.259\\
	$6/17$  & 0.816 & -0.422 & 1.238\\
	$6/18$  & 0.720 & -0.495 & 1.215 
	\end{tabular}
\end{table}	

Note that this code does not have {\em any} parallel edges in the LT code part. 
The initial code rate, or the precode code rate, is $3/4$. The subsequent code rates $6/9, 6/10, \dots, 6/18$ are obtained by transmitting the variable nodes of the LT code part starting from the top. We observe an increase in the gap between the threshold and the capacity as the code rate decreases. This is due to the structural restrictions imposed on the protograph of the LT code part.  Each subsequent protograph inherits the connections of the next-higher-rate protograph;  the new protograph can only optimize over the connections emanating from the one additional check node.  In addition, the new check node must connect with a new degree-one variable node.

\begin{figure}[t]
\centering
\includegraphics[width=0.36\textwidth]{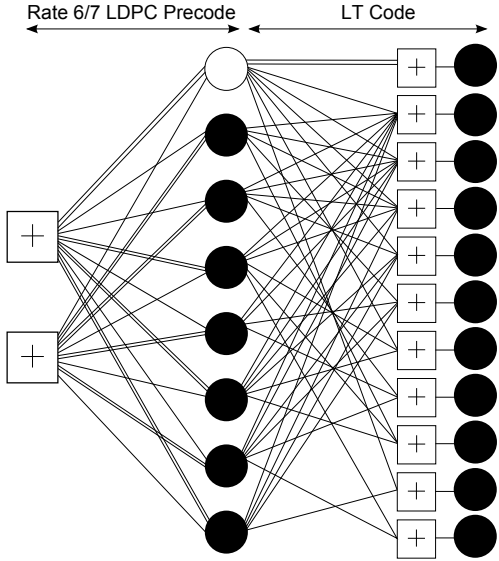}
\caption{Protograph of a PN-PBRL code with a rate-$6/7$ precode. The corresponding protomatrices are shown in \eqref{eqn:onepar1} and \eqref{eqn:onepar2}. The first node in the precode is always punctured, denoted as a white circle. Lower-rate codes are obtained by transmitting the variable nodes in the LT code protograph starting from the top node.}
\label{fig:r34_ProtoRaptor_PN-PBRL}
\end{figure}	

\subsection{Punctured-Node PBRL Codes}
\label{sec:sPN-PBRL}
As explained in  \cite{Divsalar_2009}, puncturing a node in the protograph can improve the threshold performance.  This section introduces a variant of PBRL codes for which the protograph of the precode has at least one punctured (untransmitted) node. We refer to this class of codes as the Punctured-Node PBRL (PN-PBRL) codes.

The optimization of the precode protograph is the same as described in Sec. \ref{sec:precode} except that one variable node is punctured.  
The optimization of the LT protograph is the same as in Sec. \ref{sec:LTcode} but with a slight modification  to step 2 of Algorithm \ref{OptProcedure}.   In step 2 for PN-PBRL codes, whenever there is a tie in the decoding threshold for different connections, preference is given to the one with no edge connected to the punctured node. This heuristic helps prevent performance degradation when lifting the protograph to a shorter code (for blocklengths less than $1000$).   This heuristic speeds up the decoding convergence because the punctured node has less reliability than the other variable nodes in the early iterations.

Fig.~\ref{fig:r34_ProtoRaptor_PN-PBRL} shows an example of an optimized short-blocklength PN-PBRL code. This protograph has a single parallel edge in the LT part.  Note that the first variable node of the precode protograph is punctured, giving a rate-$6/7$ precode.  The initial code rate could be $6/7$ or to obtain an initial code rate of $3/4$, the first variable node of the LT code protograph is transmitted. The precode protomatrix is give by
\begin{align}
\label{eqn:onepar1}
H_{\text{p}}=
\left[\begin{array}{cccccccc}
2 & 1 & 1 & 2 & 1 & 2 & 1 & 2 \\
1 & 2 & 2 & 1 & 2 & 1 & 2 & 1 
\end{array} \right],
\end{align}
and the LT protomatrix is given by
\begin{align}
\label{eqn:onepar2}
H_{\text{LT}} = 
\left[\begin{array}{cccccccc}
2 & 0 & 0 & 0 & 0 & 0 & 0 & 0 \\
1 & 1 & 1 & 1 & 1 & 1 & 1 & 1 \\
1 & 1 & 1 & 1 & 1 & 1 & 1 & 0 \\
1 & 1 & 1 & 1 & 1 & 0 & 0 & 1 \\
1 & 0 & 1 & 0 & 0 & 1 & 1 & 1 \\
0 & 1 & 0 & 0 & 1 & 0 & 1 & 1 \\
1 & 0 & 0 & 1 & 0 & 1 & 0 & 0 \\
0 & 0 & 1 & 0 & 1 & 0 & 1 & 0 \\
0 & 1 & 0 & 1 & 0 & 1 & 0 & 0 \\
1 & 0 & 0 & 0 & 0 & 0 & 0 & 1 \\
0 & 0 & 0 & 1 & 0 & 0 & 1 & 0 \\
\end{array} \right]
\end{align}

\begin{table}[t]  
	\caption{Thresholds of the PN-PBRL LDPC code family of Fig.~\ref{fig:r34_ProtoRaptor_PN-PBRL} with one pair of parallel edges ($E_b/N_0$ in decibels). }	
	\label{Threshold_PN-PBRL}
	\centering 
	\begin{tabular}{c|ccc} 
 	Rate & Threshold  & Capacity & Gap \\  
	\hline 
	$6/8$   & 2.020 & 1.626  & 0.394\\ 
	$6/9$   & 1.638 & 1.059  & 0.579\\
	$6/10$  & 1.468 & 0.679  & 0.789\\
	$6/11$  & 1.352 & 0.401  & 0.951\\
	$6/12$  & 1.248 & 0.187  & 1.061\\
	$6/13$  & 1.186 & 0.018 & 1.168\\
	$6/14$  & 1.018 & -0.122 & 1.140\\
	$6/15$  & 0.930 & -0.238 & 1.168\\
	$6/16$  & 0.848 & -0.337 & 1.185\\
	$6/17$  & 0.692 & -0.422 & 1.114\\
	$6/18$  & 0.602 & -0.495 & 1.097 
	\end{tabular}
\end{table}	

The subsequent code rates of $6/9, 6/10, \dots, 6/18$ are obtained by transmitting the variable nodes of the LT code protograph from top to bottom.  Regardless of the operating rate, the first variable node of the precode protograph is always punctured. The PN-PBRL codes yield better thresholds as shown in Table~\ref{Threshold_PN-PBRL} due to both the punctured node and the additional pair of parallel edges.

\subsection{Optimization for Long-Blocklength PN-PBRL Codes} \label{sec:lPN-PBRL}

For longer blocklengths, lifting can avoid short cycles even with multiple parallel edges in the LT protograph.  In fact, adding more parallel edges between the punctured variable node and the check nodes in the LT code protograph reduces the threshold significantly.  This section investigates a long-blocklength PN-PBRL optimization approach that allows multiple parallel edges in the LT protograph.

Similar to its short-blocklength counterpart, the design begins by finding a good protograph LDPC code to serve as a precode, i.e., a protograph with linear growth of its minimum distance as a function of blocklength. The construction of the LT code part, however, requires different design criteria than for the short-blocklength case in order to obtain good thresholds at each subsequent code rate. 

We observed that the punctured node of the precode must connect to all check nodes in the LT code part with at least a single edge. These edges induce high-degree punctured variable nodes at low rates.  We observe experimentally that the removal of these edges results in a notable degradation of the thresholds at low rates.  

Note the contrast between the emphasis on connecting the punctured node to all check nodes for long blocklengths and the avoidance of the punctured node in the case of ties for short blocklengths discussed in Sec.~\ref{sec:sPN-PBRL}. This contrast follows from the fact that the performance of the lifted short-blocklength codes does not always agree with the threshold results because of short cycles.

Additional pairs of parallel edges between the punctured variable node and the check nodes in the LT code part further reduce the threshold. The threshold improvement, however, is diminishing as the number of parallel edges increase.  Moreover, with numerous parallel edges connected the punctured node, the error rate performance does not match the threshold gain. In these cases, the lifted codes often display high error floors by lifting with cPEG alone. We propose two solutions to further improve the performance of the lifted code: 1) increase the size of the protograph to provide more degrees of freedom in the optimization of the lifting process and 2) optimize the lifting process with cPEG \textit{and} the Approximate Cycle Extrinsic message degree algorithm (ACE) \cite{Tian_selective_2004}. 

The construction procedure for long-blocklength PN-PBRL codes is similar to Algorithm~\ref{OptProcedure} but with some modifications for steps 1), 2) and 4). The maximum number of parallel edges connected to the puncture node is predetermined and executed in the first step. The edges in the LT part can be either single edges or parallel edges depending on the size of the protograph.
\begin{algorithm}
Given a good protograph precode, optimize the LT code part as follows:
\begin{enumerate}
	\item[1')] Add a new check node and a new degree-one variable node connected to the new check node. The new check node is connected to the punctured node with a single edge or a pair of parallel edges.
	\item[2')] Use the modified RCA to find the connections between the new check node and all the un-punctured precoded symbols that gives the lowest threshold. 
	\item[3)] If the lowest rate desired has been obtained, continue to step 4.  Otherwise, go to step 1.
	\item[4')] Lift the resulting protograph with circulant permutations to match the desired blocklengths for the rate-compatible family. The selection of the circulant permutation is based on cPEG and optionally on ACE.
	\end{enumerate}	
\end{algorithm}

The following $H_{\text{p}}$ and $H_{\text{LT}}$ gives an example of the optimized protograph with at most two pairs of parallel edges between the punctured variable node and the check nodes in the LT part. The number of pairs of parallel edges is two in this example, each of which emanates from the first two check nodes in the LT part, respectively. The precode protomatrix is given as
\begin{align}
\label{eqn:twopar1}
H_{\text{p}}=
\left[\begin{array}{cccccccc}
2 & 1 & 2 & 1 & 2 & 1 & 2 & 1 \\
1 & 2 & 1 & 2 & 1 & 2 & 1 & 2 
\end{array} \right],
\end{align}
and the LT protomatrix is given as
\begin{align}
\label{eqn:twopar2}
H_{\text{LT}} = 
\left[\begin{array}{cccccccc}
2 & 0 & 1 & 0 & 0 & 0 & 0 & 0 \\
2 & 0 & 1 & 0 & 1 & 0 & 1 & 0 \\
1 & 0 & 1 & 1 & 1 & 0 & 1 & 1 \\
1 & 0 & 1 & 0 & 1 & 0 & 1 & 1 \\
1 & 0 & 1 & 0 & 1 & 0 & 1 & 0 \\
1 & 0 & 1 & 0 & 1 & 0 & 0 & 1 \\
1 & 0 & 1 & 1 & 0 & 0 & 0 & 1 \\
1 & 0 & 1 & 0 & 1 & 0 & 1 & 1 \\
1 & 0 & 1 & 0 & 1 & 0 & 1 & 0 \\
1 & 0 & 1 & 0 & 0 & 1 & 0 & 1 \\
1 & 0 & 1 & 0 & 1 & 0 & 0 & 0 
\end{array} \right]
\end{align}
The resulting thresholds obtained from the protograph are shown in Table~\ref{Threshold_PNPBRL_Par2}. The decoding thresholds in Table~\ref{Threshold_PNPBRL_Par2} are significantly lower than the thresholds in the example in Sec.~\ref{sec:sPN-PBRL} (c.f. Table~\ref{Threshold_PN-PBRL}).  The largest gap is reduced from more than $1$ dB to less than $0.5$ dB.  All of the gaps but one are below $0.4$ dB.  This is mainly because we allow an extra pair of parallel edges and require a high-degree punctured node in the protograph. Sec.~\ref{sec:SimLongPBRL} gives an example of constructing a long-blocklength PN-PBRL code from this protograph and presents the simulation results.

\begin{table}[t]  
	\caption{Thresholds of the PN-PBRL code family shown in \eqref{eqn:twopar1} and \eqref{eqn:twopar2} with two pairs of parallel edges.  ($E_b/N_0$ in decibels). }	
	\label{Threshold_PNPBRL_Par2}
	\centering 
	\begin{tabular}{c|ccc} 
 	Rate & Threshold  & Capacity & Gap \\  
	\hline 
	$6/7$	& 3.077	& 2.625  & 0.452\\
	$6/8$   & 1.956 & 1.626  & 0.330\\ 
	$6/9$   & 1.392 & 1.059  & 0.333\\
	$6/10$  & 1.078 & 0.679  & 0.399\\
	$6/11$  & 0.798 & 0.401  & 0.397\\
	$6/12$  & 0.484 & 0.187  & 0.297\\
	$6/13$  & 0.338 & 0.018  & 0.320\\
	$6/14$  & 0.144 & -0.122 & 0.266\\
	$6/15$  & 0.072 & -0.238 & 0.310\\
	$6/16$  & 0.030 & -0.337 & 0.367\\
	$6/17$  & -0.024 & -0.422 & 0.398\\
	$6/18$  & -0.150 & -0.495 & 0.345 
	\end{tabular}
\end{table}

\begin{table}[t]  
	\caption{Thresholds of the PN-PBRL code family with parallel edges connected between the punctured node and the odd check nodes in the LT part ($E_b/N_0$ in decibels). }	
	\label{Threshold_PNPBRL_Eg3}
	\centering 
	\begin{tabular}{c|ccc} 
 	Rate & Threshold  & Capacity & Gap \\  
	\hline 
	$8/10$  & 2.179  & 2.040  & 0.139 \\
	$8/11$  & 1.579  & 1.459  & 0.120 \\ 
	$8/12$  & 1.199  & 1.059  & 0.140 \\
	$8/13$  & 0.897  & 0.762  & 0.135 \\
	$8/14$  & 0.669  & 0.530  & 0.139 \\
	$8/15$  & 0.462  & 0.342  & 0.120 \\
	$8/16$  & 0.308  & 0.187  & 0.121 \\
	$8/17$  & 0.173  & 0.056  & 0.117 \\
	$8/18$  & 0.072  & -0.056 & 0.128 \\
	$8/19$  & -0.018 & -0.153 & 0.135 \\
	$8/20$  & -0.102 & -0.238 & 0.136 \\
	$8/21$  & -0.174 & -0.314 & 0.140 \\
	$8/22$  & -0.236 & -0.381 & 0.145 \\
	$8/23$  & -0.292 & -0.441 & 0.149 \\
	$8/24$  & -0.340 & -0.495 & 0.155 \\
	$8/25$  & -0.384 & -0.545 & 0.161 \\
	$8/26$  & -0.447 & -0.590 & 0.143 \\
	$8/27$  & -0.488 & -0.631 & 0.143 \\
	$8/28$  & -0.520 & -0.669 & 0.149 \\
	$8/29$  & -0.557 & -0.704 & 0.147 \\
	$8/30$  & -0.582 & -0.736 & 0.154 \\
	$8/31$  & -0.607 & -0.766 & 0.159 \\
	$8/32$  & -0.630 & -0.794 & 0.164 \\
	\end{tabular}
	\vskip-1em
\end{table}


\setcounter{equation}{15} 
\begin{figure*}
\begin{equation}
	\label{eqn:PBRL_PrecodeCirculant}
	H_{\text{p}}^{(1)}=\left[
	\begin{array}{cccccccc}
	\sigma^0+\sigma^1+\sigma^3+\sigma^7 & \sigma^{24} & \sigma^{14} & \sigma^{17} + \sigma^0 & \sigma^7 & \sigma^1 + \sigma^6 & \sigma^{21} & \sigma^{21} + \sigma^0 \\
	\sigma^4  &  \sigma^4 + \sigma^9 & \sigma^0 + \sigma^1 & \sigma^{0} & \sigma^0 + \sigma^2 &\sigma^0 & \sigma^0 + \sigma^3 & \sigma^2
	\end{array}\right]\,.
\end{equation}
\hrulefill
\vskip-0.6em
\end{figure*}
\normalsize



\setcounter{equation}{17}
\begin{figure*}
\begin{equation}
\label{eqn:sPNPBRL_PrecodeCirculant}
	H_{\text{p}}^{(2)}=\left[\begin{array}{cccccccc}
	\sigma^0+\sigma^1 & \sigma^{24} & \sigma^{14} & \sigma^{17} + \sigma^5 & \sigma^7 & \sigma^1 + \sigma^3 & \sigma^{21} & \sigma^{21} + \sigma^0 \\
	\sigma^4  &  \sigma^0 + \sigma^2 & \sigma^0 + \sigma^3 & \sigma^{31} & \sigma^6 + \sigma^0 &\sigma^1 & \sigma^0 + \sigma^1 & \sigma^2
	\end{array} \right]\,.
\end{equation}
\hrulefill
\vskip-1.5em
\end{figure*}
\setcounter{equation}{18}
\normalsize

By adding parallel edges between the punctured node and {\em all} the check nodes in the LT part, the gaps between the thresholds and the capacities are improved even more than shown in Table~\ref{Threshold_PN-PBRL} to less than $0.34$ dB except for the highest rate (the precode).  However, the lifted codes using cPEG do not lead to good error rate performance.  A larger protograph is needed to obtain better thresholds and have a controlled number of parallel edges connected to the punctured variable node. 

We constructed a protograph with $33$ variable nodes.  The protomatrix is omitted due to page limitations and can be found on the CSL website \cite{CSLPAGE}. Parallel edges are added heuristically to all the odd check nodes in the LT part and the selection of other connections follows the greedy RCA  optimization described earlier. The connections between the check nodes in the LT part and the un-punctured variable nodes can be single edge or parallel edges, i.e., the matrix entries can be $0$, $1$ or $2$. The resulting thresholds are summarized in Table~\ref{Threshold_PNPBRL_Eg3}. The thresholds are significantly lower than the thresholds in Table~\ref{Threshold_PNPBRL_Par2} and the gaps to capacities are all within $0.164$ dB.

\section{Examples and Simulations of PBRL Codes}
\label{sec:ExamplesPBRL}
This section provides examples of PBRL codes in the short-blocklength regime ($k = 192$) and the long-blocklength regime ($k = 16368$ and $16384$). Simulation results of PBRL codes are presented and compared to other channel codes with similar blocklengths in the literature. All of our simulations use iterative decoders with floating-point computation, and the schedule of the message-passing algorithm is flooding unless otherwise stated. The iterative decoder terminates if the decoding is successful (passing the parity check) before reaching the maximum number of iterations. The maximum number of iterations is $100$ for all simulations unless otherwise stated.

\subsection{Short-Blocklength PBRL Codes}
\label{sec:SimShortPBRL}
This subsection provides examples of PBRL and PN-PBRL codes with short-blocklengths. Lifting of the protograph is accomplished by circulant permutation of the edges of each type, which allows efficient implementation of the decoder. The design of the circulant permutation uses cPEG algorithm. For the following short-blocklength examples, cPEG avoided all length-$4$ cycles and minimized the number of length-$6$ cycles. 
Experimental results indicate that for PBRL codes with $k = 192$, direct lifting of the protograph yields better codes than the two-stage lifting such described in \cite{Divsalar_2009}.

\begin{figure}[t]
\centering
\includegraphics[width=0.45\textwidth]{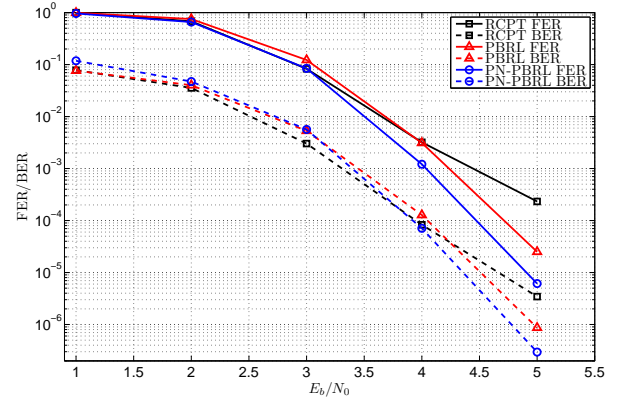}
\caption{Frame error rate and bit error rate of the PBRL code, PN-PBRL code and RCPT code at code rate $3/4$. Both PBRL and PN-PBRL codes outperform the RCPT codes at high $E_b/N_0$ regime but perform slightly worse than the RCPT code in the low $E_b/N_0$ regime.}
\label{fig:FER_HighRate}
\vskip-0.5em
\end{figure}

\begin{figure}[t]
\centering
\includegraphics[width=0.45\textwidth]{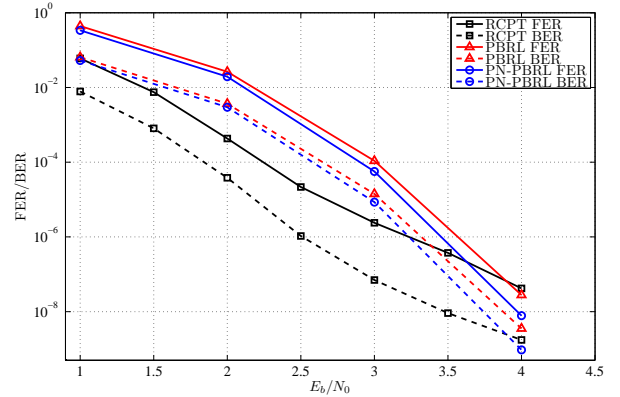}
\caption{Frame error rate and bit error rate of the PBRL code, PN-PBRL code and RCPT code at code rate $1/3$. Here the RCPT code outperforms the PN-PBRL and PBRL code at low SNR range, but the PN-PBRL code starts to outperform the RCPT code at around SNR $3.5$ dB. There is no sign of an error floor for the both PBRL and PN-PBRL code.}
\label{fig:FER_LowRate}
\vskip-0.5em
\end{figure}	

\setcounter{equation}{19}
\begin{figure*}[t]

\begin{equation}
	\label{eqn:lPNPBRL_PrecodeCirculant}
	H_{\text{p}}^{(3)} =\left[\begin{array}{cccccccc}
	\sigma^0+\sigma^1 & \sigma^3 & \sigma^1+\sigma^0 & \sigma^2 & \sigma^0 + \sigma^1 & \sigma^0 & \sigma^2 + \sigma^0  & \sigma^2 \\
	\sigma^0  &  \sigma^0 + \sigma^1 & \sigma^0 & \sigma^0 + \sigma^1 & \sigma^0 &\sigma^0 + \sigma^1 & \sigma^0 & \sigma^0 + \sigma^1
	\end{array} \right].
\end{equation}
\normalsize
\hrulefill
\vskip-1em
\end{figure*}
\setcounter{equation}{20}

Recall from \eqref{eqn:protomatrix_r34_eg} that it is enough to specify $H_{\text{p}}$ and $H_{\text{LT}}$ to express the protomatrix of a PBRL code. Let $\sigma$ be a $32\times 32$ identity matrix shifted to the left by $1$, and let $0$ be a $32\times 32$ all-zero matrix. We index the first example by the superscript $(1)$. The lifted precode part $H_{\text{p}}^{(1)}$ is given by \eqref{eqn:PBRL_PrecodeCirculant}, shown at the top of the page, and the LT part $H_{\text{LT}}^{(1)}$ is given by
\setcounter{equation}{16}
\begin{equation}
\makeatletter
    \def\tagform@#1{\maketag@@@{\normalsize(#1)\@@italiccorr}}
\makeatother
\footnotesize
	\label{eqn:PBRL_LTCirculant}
	H_{\text{LT}}^{(1)} =\left[\begin{array}{cccccccc}
	 \sigma^{29} & \sigma^0 & \sigma^0  & \sigma^1  & \sigma^5 & \sigma^6 & \sigma^{10} & \sigma^4    \\ 
	 \sigma^{12} & \sigma^0 & \sigma^1  & \sigma^3  & \sigma^4 & \sigma^{16} & \sigma^{13} & 0    \\ 
	 \sigma^{16} & \sigma^0 & \sigma^2  & \sigma^6  & \sigma^0 & 0   & 0   & \sigma^{1}  \\ 
	 \sigma^{26} & 0   & \sigma^0  & 0    & 0   & \sigma^1 & \sigma^6 & \sigma^9     \\  
	 0      & \sigma^1 & 0    & 0    & \sigma^0 & 0   & \sigma^2 & \sigma^3    \\ 
	 \sigma^1    & 0   & 0    & \sigma^2  & 0   & \sigma^9 & 0   & 0      \\ 
	 0      & 0   & \sigma^{16}& 0   & \sigma^0 & 0   & \sigma^4 & 0     \\ 
	 0      & \sigma^{21}& 0  & \sigma^0  & 0   & \sigma^2 & 0   & 0       \\ 
	 \sigma^0    & 0   & 0    & 0    & 0   & 0   & 0   & \sigma^1      \\ 
	 0      & 0   & 0    & \sigma^{12}& 0  & 0   & \sigma^0 & 0       \\ 
	\end{array} \right],
\end{equation}
\normalsize
where entries with multiple terms of $\sigma$ in \eqref{eqn:PBRL_PrecodeCirculant} indicate parallel edges in the protograph.

For the PN-PBRL code example the lifted precode $H_{\text{p}}^{(2)}$ is given by \eqref{eqn:sPNPBRL_PrecodeCirculant}, shown at the top of the page, and the lifted LT code $H_{\text{LT}}^{(2)}$ is given by
\setcounter{equation}{18}
\begin{equation}
H_{\text{LT}}^{(2)}=
\makeatletter
\def\tagform@#1{\maketag@@@{\normalsize(#1)\@@italiccorr}}
\makeatother
\footnotesize
\label{eqn:sPNPBRL_LTCirculant}
	\left[\begin{array}{cccccccc}
	 \sigma^2+\sigma^0 & 0  & 0    & 0    & 0   & 0   & 0   & 0        \\ 
	 \sigma^{29} & \sigma^0 & \sigma^2  & \sigma^0  & \sigma^9 & \sigma^6 & \sigma^7 & \sigma^6    \\ 
	 \sigma^{12} & \sigma^0 & \sigma^4  & \sigma^1  & \sigma^5 & \sigma^4 & \sigma^{10} & 0    \\ 
	 \sigma^{16} & \sigma^0 & \sigma^5  & \sigma^6  & \sigma^1 & 0   & 0   & \sigma^{11}  \\ 
	 \sigma^{26} & 0   & \sigma^0  & 0    & 0   & \sigma^2 & \sigma^9 & \sigma^0     \\  
	 0      & \sigma^1 & 0    & 0    & \sigma^0 & 0   & \sigma^3 & \sigma^0    \\ 
	 \sigma^1    & 0   & 0    & \sigma^0  & 0   & \sigma^0 & 0   & 0       \\ 
	 0      & 0   & \sigma^{16}& 0   & \sigma^0 & 0   & \sigma^2 & \sigma^0     \\ 
	 0      & \sigma^{21}& 0  & \sigma^0  & 0   & \sigma^1 & 0   & 0       \\ 
	 \sigma^0    & 0   & 0    & 0    & 0   & 0   & 0   & \sigma^1      \\ 
	 0      & 0   & 0    & \sigma^{12}& 0  & 0   & \sigma^0 & 0       \\ 
	\end{array} \right].
\end{equation}
\normalsize

Note that the matrix in \eqref{eqn:PBRL_LTCirculant} has no parallel edge The matrix in \eqref{eqn:sPNPBRL_LTCirculant} corresponds to the LT protograph in Fig.~\ref{fig:r34_ProtoRaptor_PN-PBRL}, which has one pair of parallel edges represented as $\sigma^2+\sigma^0$.

For ease of comparison, Figs.~\ref{fig:FER_HighRate} and \ref{fig:FER_LowRate} separately plot rates $3/4$ and $1/3$ for the PBRL code family, PN-PBRL code family, and the RCPT codes in \cite{3GPP} showing both frame error rate (FER) and bit error rate (BER). Flooding is used for the decoder simulations. Consistent with the decoding threshold results, the PN-PBRL code family outperforms the PBRL code family with a slight increase of encoding complexity due to the punctured node. 

At rate $3/4$, the PN-PBRL code performs similarly to the RCPT code and outperforms the RCPT code when SNR is higher than $3$ dB in terms of FER and 4dB in terms of BER. At rate $1/3$, the PN-PBRL code starts to gain an advantage at SNR higher than $3.5$ dB in terms of FER and 4 dB for BER. 

We omit the simulation results for short-blocklength PBRL codes with rates between $1/3$ and $3/4$ due to page limitations. We refer readers to \cite{Chen_2011_Globecom} and \cite{Chen_dissertation_2013} for more simulation results.  In all cases, the PN-PBRL code family outperforms the RCPT code family at a sufficiently high SNR, and the crossover where PN-PBRL outperforms RCPT occurs at higher FER/BER for higher rates.

Simulation results using Raptor codes with the same precode $H_{\text{p}}^{(1)}$ as the PBRL code family with output distributions drawn from \cite{Etesami_2006} and \cite{Soljanin_2006} show a much higher FER in BI-AWGN with information blocklength $192$ than either PBRL or PN-PBRL codes at all rates. Note that the comparison even ignores the overhead for communicating randomly drawn LT connections. The result is not surprising because the Raptor codes do not transmit the precode symbols and the degrees of each output node are drawn at random according to the optimal degree distribution, and a few hundreds of samples may not be enough to exhibit the optimal degree distribution. 


\subsection{Long-Blocklength PBRL Codes}
\label{sec:SimLongPBRL}
This subsection provides the simulation results of the two families of long-blocklength PN-PBRL codes presented in Sec.~\ref{sec:lPN-PBRL}. The lifting process is discussed in detail using the protomatrices \eqref{eqn:twopar1} and \eqref{eqn:twopar2}. We omit the details for lifting the protograph with $33$ variable nodes due to page limitations. The code can be found on the CSL website \cite{CSLPAGE}.

As in \cite{Divsalar_2009} we use two-stage lifting to obtain good protograph codes with long blocklengths. The first stage, also known as pre-lifting, uses a relatively small lifting number (i.e. the number of replicas) and aims to remove the parallel edges in the protograph. The second stage then lifts the protograph resulting from the previous stage to the desired blocklength. In this example, the protograph based on \eqref{eqn:twopar1} and \eqref{eqn:twopar2} is pre-lifted $4$ times and then further lifted $682$ times, the resulting information blocklength is then $k = 16368$. 

With a step size of $4\times 682=2728$, subsequent code rates $6/8, 6/9, \dots, 6/18$ are obtained by transmitting the output symbols of the LT code from each successive protograph node starting from the top.  However, many more rates between  $1/3$ and $6/7$ can be obtained by adding one variable node at a time to the lifted graph. 

Similar to the short-blocklength codes, lifting of the protograph is also accomplished by circulant permutation and the design of the circulant permutation uses the cPEG algorithm. The minimum cycle in the lifted graph of the precode has length $10$ while the minimum cycle in the overall lifted graph of the LT and precode  at the lowest rate of $1/3$ has length $8$.

\setcounter{equation}{20}

As with the short-blocklength codes,  the parity-check matrices of the pre-lifted code and the final lifted code both have the structure of \eqref{eqn:protomatrix_r34_eg}.  The pre-lifted  precode $H_{\text{p}}^{(3)}$ is given in \eqref{eqn:lPNPBRL_PrecodeCirculant}, shown at the top of the page with $\sigma$ a $4\times 4$ identity matrix shifted to the left by $1$ and $0$ representing the $4\times 4$ all-zero matrix.   Similarly, $H_{\text{LT}}^{(3)}$ is given as
\begin{equation}
\makeatletter
    \def\tagform@#1{\maketag@@@{\normalsize(#1)\@@italiccorr}}
\makeatother
\footnotesize
\label{eqn:lPNPBRL_LTCirculant}
	H_{\text{LT}}^{(3)} =\left[
	\begin{array}{cccccccc}
	 \sigma^3+\sigma^0 & 0 & \sigma^0  & 0    & 0   & 0   & 0   & 0        \\ 
	 \sigma^1+\sigma^0 & 0 & \sigma^0  & 0    & \sigma^0 & 0 & \sigma^0 & 0    \\ 
	 \sigma^2          & 0 & \sigma^0  & \sigma^0  & \sigma^0 & 0 & \sigma^0 & \sigma^0\\ 
	 \sigma^3 & 0  & \sigma^0  & 0  & \sigma^0 & 0 & \sigma^0 & \sigma^0 \\ 
	 \sigma^0 & 0  & \sigma^0  & 0  & \sigma^0 & 0 & \sigma^0 & 0     \\  
	 \sigma^0 & 0  & \sigma^0  & 0  & \sigma^0 & 0 & 0 & \sigma^0     \\ 
	 \sigma^2 & 0  & \sigma^0  & 0  & \sigma^0 & 0 & \sigma^0 & \sigma^2       \\ 
	 \sigma^0 & 0  & \sigma^0  & \sigma^0  & 0 & 0 & 0 & \sigma^0     \\ 
	 \sigma^0 & 0  & \sigma^0  & 0  & \sigma^0 & 0 & \sigma^0 & 0     \\ 
	 \sigma^1 & 0  & \sigma^0  & 0  & 0 & \sigma^0  & 0   & \sigma^0      \\ 
	 \sigma^0 & 0  & \sigma^0  & 0  & \sigma^0  & 0 & 0   & 0       \\ 
	\end{array} \right].
\end{equation}
\normalsize
\begin{figure}[t]
\centering
	\includegraphics[width = 0.5\textwidth]{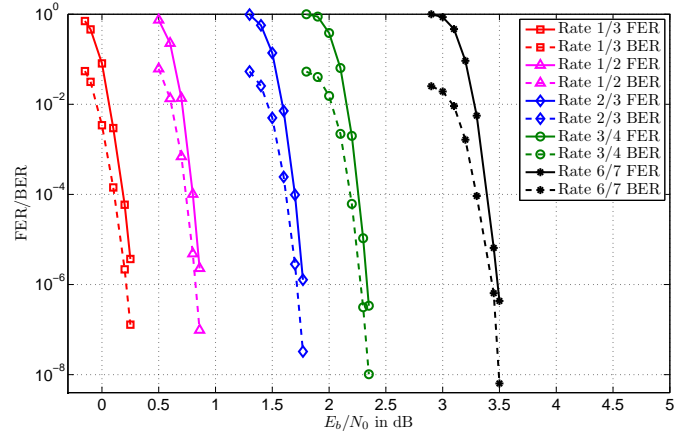}
	\caption{Frame error rates and bit error rates for the protograph based on \eqref{eqn:twopar1} and \eqref{eqn:twopar2}. The lifting number is $2728$, resulting in $k = 16368$.}
	\label{fig:FER_PNPBRL}
\end{figure}

The pre-lifted protograph contains no parallel edges and is further lifted by $682$ using powers of a larger circulant matrix $\Sigma$: a $682\times 682$ identity matrix shifted to the left by $1$. The powers of $\Sigma$, i.e. the assignments of the circulant for all edges in the pre-lifted protograph, are omitted due to space limit and but are available in \cite{CSLPAGE}.

Fig.~\ref{fig:FER_PNPBRL} shows the simulations of the example PN-PBRL code family. The simulated codes have rates $6/7, 3/4, 2/3, 1/2$ and $1/3$ with the corresponding blocklengths $19096$, $21824$, $24552$, $32736$ and $49104$, respectively.  All other rates in the form of $6/m, m = 8, 9, \dots, 17$ are also simulated with similar performance in the waterfall region and no error floor has been observed for FER greater than $10^{-6}$ \cite{Chen_dissertation_2013}. 

Table~\ref{table:GapToCapacity} presents the gaps between Shannon limits and the SNRs required to achieve a fixed FER of $10^{-5}$ at various rates. Only the rates presented in Fig.~\ref{fig:FER_PNPBRL} are shown due to space limit, but the gaps for all other rates range from $0.643$ dB (at rate $1/2$) to $0.765$ dB (at rate $6/7$).

\begin{table}[t]  
	\caption{SNRs Required to Achieve FER $10^{-5}$ for for the first long-blocklength PN-PBRL code family. }	
	\label{table:GapToCapacity}
	\centering 
	\begin{tabular}{c|ccc} 
 	Rate & Req. SNR & Shannon limit & Gap \\  
	\hline 
	$6/7$  & 3.39 & 2.625    & 0.765 \\ 
	$6/8$   & 2.30 & 1.626   & 0.674 \\
	$6/9$   & 1.74 & 1.059   & 0.681 \\
	$6/12$  & 0.83 & 0.187   & 0.643 \\
	$6/18$  & 0.23 & -0.495  & 0.725 
	\end{tabular}
\end{table}

\begin{figure}[t]
\centering
	\includegraphics[width = 0.5\textwidth]{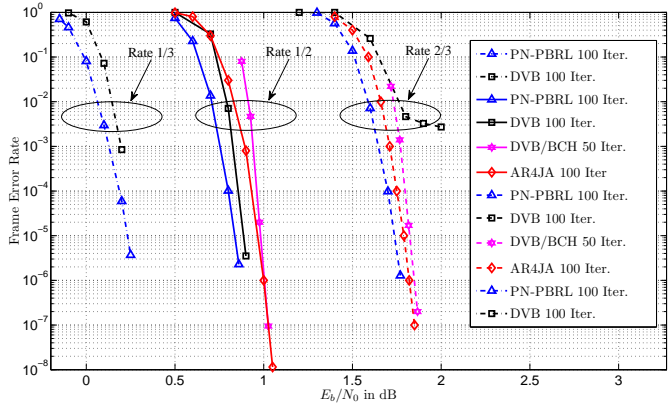}
	\caption{Frame error rates comparison for PN-PBRL codes, LDPC codes in the DVB-S2 standard, and AR4JA LDPC codes in the CCSDS standard.}
	\label{fig:DVBS2_AR4JA_PBRL}
\end{figure}

Fig.~\ref{fig:DVBS2_AR4JA_PBRL} compares  the FER performance of the example PN-PBRL code family at rates $1/2$ and $2/3$ to the DVB-S2 standard (with and without an outer BCH code) \cite{DVBS2} and AR4JA codes from the CCSDS standard \cite{CCSDS_ORANGE}. In some cases for DVB-S2 codes, only results for a maximum of $50$ iterations were available.  The blocklengths of the DVB-S2 codes are fixed to $64800$ bits, whereas the PN-PBRL and AR4JA codes have a fixed information length of $16368$ bits and blocklengths of $32736$ bits and $24552$ bits for rate $1/2$ and rate $2/3$, respectively. When concatenated with the BCH code, the overall rates of DVB-S2 codes are $0.497$ bits and $0.664$ bits.

Fig.~\ref{fig:DVBS2_AR4JA_PBRL} shows that in the waterfall region, the PN-PBRL codes outperform both the AR4JA codes and the DVB-S2 codes.  Note that the PN-PBRL codes outperform DVB-S2 even though the DVB-S2 codes have longer blocklength and benefit from concatenation with a BCH code.

\begin{figure}[t]
\centering
	\includegraphics[width = 0.5\textwidth]{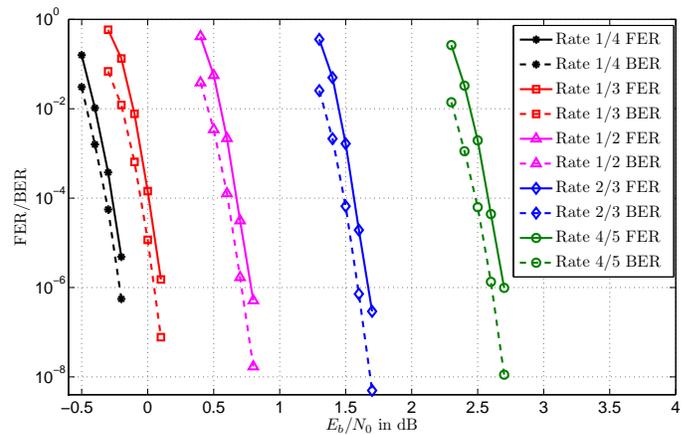}
	\caption{Frame error rates and bit error rates for the V$33$ protograph. The lifting number is $2048$, resulting in $k = 16384$. }
	\label{fig:PBRL_v33}
\end{figure}

\begin{figure}[t]
\centering
	\includegraphics[width = 0.5\textwidth]{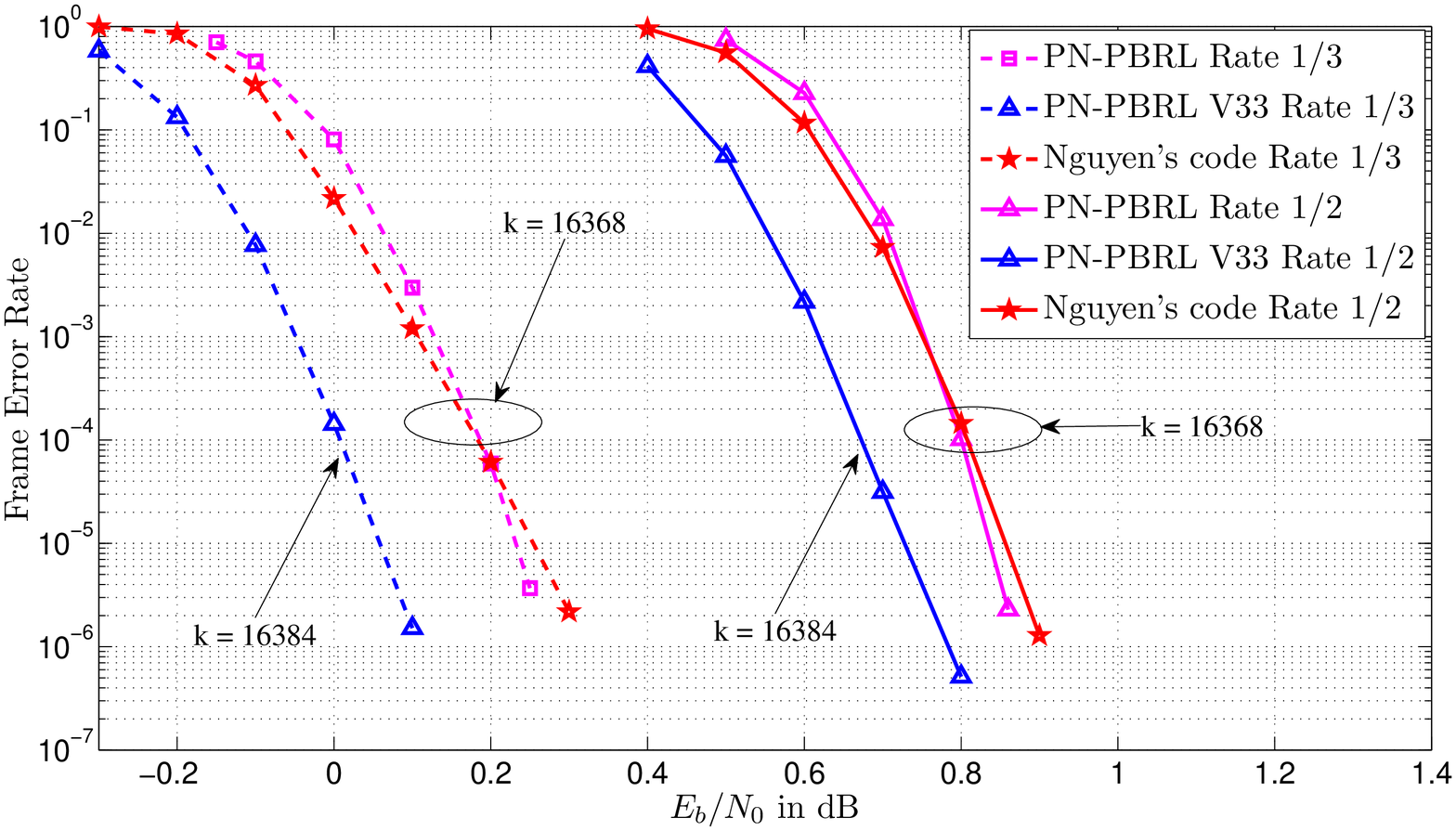}
	\caption{Frame error rates comparison between the PN-PBRL codes with $19$ variable nodes in the protograph ($k = 16368$), the PN-PBRL codes with $33$ variable nodes in the protograph ($k = 16384$), and the rate-compatible protograph codes proposed in \cite{Nguyen_design_2012} ($k = 16368$).}
	\label{fig:PBRL_Nguyen}
\end{figure}

Nguyen et al. \cite{Nguyen_design_2012} constructed a rate-compatible protograph of $47$ variables  with lower thresholds than the thresholds in Table~\ref{Threshold_PNPBRL_Par2}. As shown in Table~\ref{Threshold_PNPBRL_Eg3}, we demonstrated an example of a protograph with $33$ variable nodes and the Raptor-like structure that gives threshold comparable to those obtained in \cite{Nguyen_design_2012}. Moreover, we demonstrate in the following that the lifted code can outperform existing codes designed without the Raptor-like structure constraint.

For the protograph with $33$ variable nodes (V$33$), the lifting numbers in the first and second stages are $4$ and $512$, respectively. Both cPEG and ACE are used jointly to lift the code in the second stage. The lifting process first chooses a permutation randomly for an edge and the permutation will be admitted if it passes both cPEG and ACE constraints. A girth of $8$ is strictly enforced for cPEG, and the constraint for ACE is enforced by having $(d, \eta)$ to be $(7, 21)$ initially. If the search fails after $100$ attempts, the parameters are relaxed (still enforcing the cPEG constraint) to the ACE constraint of $(6, 21)$. The algorithm relaxes the constraint to $(6, 21)$ after $100$ attempts. The constraints are further relaxed to $(6, 20)$, $(6,19)$ and $(6,18)$ incrementally after every additional $100$ attempts. If the ACE parameters of $(6,18)$ cannot be met for an edge, the optimization will stop and the process must restart from the first edge.

Fig~\ref{fig:PBRL_v33} presents the error rate performance of various rates and Fig.~\ref{fig:PBRL_Nguyen} compares the FER between the PN-PBRL codes ($k = 16384$) and the rate-compatible protograph codes proposed in \cite{Nguyen_design_2012} ($k = 16368$) at rates $1/3$ and $1/2$. The simulation results in \cite{Nguyen_design_2012} used an $8$-bit quantized decoder with $200$ iterations whereas our simulations used floating point decoder with $100$ iterations. The simulations of the V$33$ protograph in Fig.~\ref{fig:PBRL_Nguyen} used layered belief propagation. Using flooding in our simulation results in less than $0.05$dB of degradation at $100$ iterations. Comparing at FER around $10^{-6}$, PN-PBRL code is $0.2$dB better than Nguyen's code at rate $1/3$ and about $0.1$ dB better at rate-$1/2$. It is surprising that PBRL codes can have superior performance even with the constraints of the Raptor-like structure. In addition, our protograph ($33$ variable nodes) is smaller than the one constructed in \cite{Nguyen_design_2012}.

Table~\ref{table:GapToCapacityV33} presents the gaps between Shannon limits and the SNRs required to achieve a fixed FER of $10^{-5}$ at various rates. Only the rates presented in Fig.~\ref{fig:PBRL_v33} are shown due to space limit. The gaps for the simulated rates in Fig~\ref{fig:PBRL_v33} ranges from $0.533$ dB (at rate $1/2$) to $0.645$ dB (at rate $1/3$).

\begin{table}[t]  
	\caption{SNRs Required to Achieve FER $10^{-5}$ for the V33 PBRL code. }	
	\label{table:GapToCapacityV33}
	\centering 
	\begin{tabular}{c|ccc} 
 	Rate & Req. SNR & Shannon limit  & Gap \\  
	\hline 
	$8/10$ & 2.64 &  2.040  & 0.6 \\ 
	$8/12$ & 1.62 &  1.059  & 0.561 \\
	$8/16$ & 0.72 &  0.187  & 0.533 \\
	$8/24$ & 0.15 & -0.495  & 0.645 \\
	$8/32$ & -0.22 & -0.794  & 0.574 
	\end{tabular}
\end{table}

In general, larger protographs have more degrees of freedom and hence yield lower thresholds after optimization but with diminishing returns. In addition, large protographs are less flexible in the lifting optimization, e.g. lifting a very large protograph (e.g. the V$33$ protograph) for a moderate-blocklength code (e.g. $n = 2000$) may harm the lifting optimization since there are less permutations to choose from compare to a moderate-size protograph.

In \cite{Nguyen_rate_2013}, Nguyen et al. constructed a family of rate-compatible protograph codes with moderate blocklengths that turn out to have the PBRL structure. Using EXIT analysis, \cite{Nguyen_rate_2013} proposed design principles similar to the current paper. The example in \cite{Nguyen_rate_2013} uses $7$ pairs of parallel edges in the LT protograph in contrast to the example shown in the current paper, which has at most $1$ pair of parallel edges to avoid error floors for blocklengths less than $1000$. The simulation results in \cite{Nguyen_rate_2013} for $k = 1024$ are remarkable, with no error floor even at the highest SNR studied.

The results in \cite{Nguyen_rate_2013} suggest that the design of PBRL codes may be further improved by adding even more parallel edges when a quantized decoder is used. Regardless, the outstanding numerical results in \cite{Nguyen_rate_2013} further verifies that Raptor-like structure enjoys extensive rate-compatibility, helps reduce the complexity in encoding and decoding, and provides excellent performance.

\section{Concluding Remarks}
\label{sec:ConclusionPBRL}

This paper studies the construction and optimization of protograph-based Raptor-like LDPC codes. The optimization of PBRL codes is based on asymptotic results of LDPC codes, i.e., density evolution. Instead of the original density evolution, a modified reciprocal channel approximation is used to obtain a fast and accurate approximation for the thresholds of PBRL codes. The assignment of the circulants when lifting the codes is based on cPEG algorithm and ACE algorithm.

Puncturing variable nodes in the precode protograph further improves the threshold performance of PBRL codes. This class of PBRL codes is referred to as the PN-PBRL codes. PN-PBRL codes have better performance but a slightly more complicated encoder for the initial transmission. 

For long-blocklength codes, optimization using density evolution provides useful guidance to improve the performance of the PBRL code. For short-blocklength codes, the design procedure must avoid undesirable graphical structures while minimizing the threshold. 

A threshold saturation is observed as the rate decreases in the optimization of PBRL codes. Adding more pairs of parallel edges in the LT protograph alleviates the saturation issue and yields low thresholds. The lifted codes with many pairs parallel edges, however, do not always perform better than the codes with fewer pairs of parallel edges. For the rates considered in this paper, we observed that one pair of parallel edges gives the best result for short-blocklength codes ($k = 192$). For long-blocklength codes ($k = 16368$ and $16384$), we observed that two pairs of parallel edges give excellent performance for small protographs, while for multiple pairs of parallel edges, optimizing over a larger protograph and using a better lifting algorithm are necessary. Specifically, we used cPEG and ACE jointly to obtain good codes. 

Simulation results of short-blocklength and long-blocklength PBRL codes are presented and compared to the other rate-compatible channel codes in the literature. The short-blocklength PBRL codes outperform the RCPT codes in the 3GPP-LTE standard \cite{3GPP} at high SNRs and do not have error floors up to the highest SNRs studied. The short-blocklength PBRL codes perform worse than the RCPT codes at low SNRs. The long-blocklength codes using the small protograph outperform AR4JA codes and DVB-S2 codes and do not have error floors down to FER as low as $10^{-7}$.

To compare the PBRL codes with the rate-compatible LDPC codes in \cite{Nguyen_design_2012}, we constructed a protograph that is smaller than the one considered in \cite{Nguyen_design_2012}. Although the thresholds of PBRL codes obtained by RCA are larger than the thresholds reported in \cite{Nguyen_design_2012}, the lifted codes outperform the rate-compatible LDPC codes in \cite{Nguyen_design_2012}.

\normalsize

\bibliographystyle{IEEEtran}
\bibliography{IEEEabrv,PBRL_Journal}

\end{document}